# Silver Nanoplates: Theoretical and Experimental Characterization. Study of SERS Application

Cristian Villa-Pérez,[a,b] Luis J. Mendoza-Herrera,[a,c] Jesica M. J. Santillán,[a,c] David Muñetón Arboleda,[a,c] M. Sergio Moreno,[d] Valeria B. Arce [a,b,*]

[a] Centro de Investigaciones Ópticas (CIOp), (CONICET-CIC-UNLP), Camino Centenario y 506, 1897 Gonnet, La Plata, Argentina.

[b] Departamento de Química, Facultad de Ciencias Exactas, UNLP, 115 y 49, 1900 La Plata, Buenos Aires, Argentina.

[c] Departamento de Ciencias Básicas, Facultad de Ingeniería, UNLP, 115 y 49, 1900 La Plata, Buenos Aires, Argentina.

[d] Instituto de Nanociencia y Nanotecnología, INN (CNEA-CONICET), Centro Atómico Bariloche, Av. Bustillo 9500, San Carlos de Bariloche 8400, Argentina.

*Corresponding author: E-mail address: varce@ciop.unlp.edu.ar (V.B. Arce).

**Abstract**

Metal nanostructures have received significant attention in recent years owing to their peculiar physical and chemical properties, which span a wide range of practical applications in a variety of scientific and engineering fields. Herein, we report a study of the optical properties and application in the field enhancement of silver nanoplates (Ag-Nplates) obtained by two-step chemical synthesis. Optical Extinction Spectroscopy (OES), together with a detailed theoretical analysis of the experimental spectra, was used to characterize the size distribution, morphology, and optical spectral behavior of the obtained suspensions. These studies were complemented with shape and size distribution analyses using Transmission Electron Microscopy (TEM). It was observed by both techniques that these characteristics are related to the concentration of silver spherical nanoparticles (Ag-Seeds) added during the synthesis of Ag-Nplates. A decrease in the Ag-Seed amount produces an infrared shift of the main plasmon peak with respect to the neat Ag-Seeds suspension, which is related to the size growth of Ag-Nplates showing size distributions centered in the edge length range of 12–15 nm and 34–36 nm for the experimental spectra, whose plasmon peak positions are at 543 nm (higher Ag-Seeds concentration) and 815 nm (lower Ag-Seed concentration), respectively. The detection of Brilliant Blue (BB) by Surface-Enhanced Raman Spectroscopy (SERS) was successfully probed with the incorporation of Ag-Nplates in an agarose gel.

**Introduction**

Over the last few decades, nanostructure fabrication has been extensively studied because of its unique properties compared to bulk dimensions, which allow them to be highly effective in a wide range of applications, including catalysis [1], photonics [2], electronics [3, 4], optoelectronics [5], information storage [6, 7], sensing [8], imaging [9], medical treatment [10, 11], and surface-enhanced spectroscopy [12].

A strong relationship exists between the properties of metal nanostructures and their dimensions, composition, crystallinity, shape, and construction geometry (e.g., core-shell, solid, and hollow) [13-15]. Recently, attention has been given to shape-controlled synthesis because it allows fine-tuning of the properties of nanostructures by varying their shape. Nanorods with high aspect ratios, for example, exhibit highly sensitive optical responses (e.g., Localized Surface Plasmon Resonance (LSPR) and photoluminescence) [16]. According to both theoretical calculations and experimental studies, the number of absorption peaks and effective spectral ranges of Surface-Enhanced Raman Spectroscopy (SERS) are strongly influenced by the precise morphology of metal nanostructures [17]. Nanostructures with nonspherical shapes, such as cubes [18, 19], rings [20], rods or wires [21], stars [22], and plates [23], have been fabricated using different metals and alloys. This has led to the synthesis of nanoplates (Nplates); triangular Nplates made of noble metals have become increasingly popular in recent years [24-26].

Metal Nplates are primarily fabricated using solution-phase chemical reactions with various surfactant molecules as aids, including polymeric chains (such as PVP or polyamines) [27], coordinating molecules [28], and biological reagents [29]. As a result of these approaches, plate-like and triangle-like nanostructures with smooth surfaces have been successfully prepared [30].

In particular, silver triangular Nplates are emerging as notable structures, capturing the attention of researchers and scientists owing to their unique plasmonic properties and potential in a variety of applications [31-33]. These nanostructures, characterized by their sharp edges and triangular shapes, exhibit extraordinary optical properties that make them particularly suitable for the enhancement of Raman spectroscopy through SERS [34-36]. Raman spectroscopy is a powerful analytical technique that provides detailed information on the molecular vibrations of materials, thereby allowing the identification and characterization of chemical compounds. However, their often-limited sensitivity can be overcome by the use of plasmonic substrates, such as silver triangular Nplates, which significantly amplify the Raman signal. For these reasons, the synthesis of silver triangular Nplates has become an area of intense research, with approaches ranging from traditional chemical methods [37-39] to advanced manufacturing techniques such as electron beam lithography [40]. These methods allow the size, shape, and distribution of triangular Nplates to be carefully controlled, which in turn influences their optical properties and performance in specific SERS applications, thereby offering significant potential in a variety of areas. For example, in medicine, they are explored for the early detection of diseases by identifying specific biomarkers in biological samples [41-44]. In materials science, they are used for the analysis of thin films and nanostructures, whereas in the detection of environmental contaminants, they can assist in the precise monitoring of toxic substances in water and air [45-47]. Furthermore, these nanostructures have also been applied in the fabrication of advanced optical devices, such as biosensors [48], gas sensing devices [49], and heterogeneous catalysts [50]. Their ability to improve the sensitivity and selectivity of these devices makes them valuable components for the creation of more efficient and precise technologies. In summary, silver triangular Nplates represent an exciting class of nanostructures with promising applications in a variety of fields including medicine, materials science, and environmental technology. As we continue to explore and better understand its properties and potential, its impact on research and innovation is likely to grow, opening new frontiers in nanotechnology and analytical spectroscopy. In this framework, this work deals with the synthesis and characterization of silver nanoplates (Ag-Nplates) using a two-step chemical synthesis to analyze their optical properties and field enhancement because of their great potential for application in different areas of science and technology. The size distribution, morphology, spectral characteristics, and relationship of the amount of silver seeds (Ag-Seeds) added in the synthesis with respect to the size growth of the obtained Ag-Nplate were studied using Optical Extinction Spectroscopy (OES). A theoretical analysis of the experimental spectra based on the Mie theory for metallic nanospheres (Nspheres) and on the Discrete Dipole Approximation (DDA) for Nplates was performed to obtain more detailed information about the samples. The shape and size distribution were analyzed by Transmission Electron Microscopy (TEM). The SERS effect of Ag-Nplate on the detection of Brilliant Blue (BB) was studied using Raman spectroscopy.

**Materials and Methods**

**Chemicals**

All reagents used were of analytical grade. $AgNO_3$ was purchased from Merck, polyvinylpyrrolidone, trisodium citrate, ascorbic acid, and $NaBH_4$ were obtained from Biopack. Brilliant blue FCF dye. Ultrapure water (14.3 MΩ cm) was employed for all the experiments. All chemicals were used as received, without further treatment.

**Instrumentation**

UV-Vis measurements were acquired using a Shimadzu UV1650 spectrophotometer with a 0.5 nm resolution. The generated Ag suspensions were placed in 1 cm pathlength quartz cuvettes, and spectral measurements were performed in the UV-Vis-NIR range 300–1100 nm in the as-prepared suspensions immediately after synthesis.

Raman spectra were acquired using a Jobin XploRA plus confocal Raman microscope and calibrated using a silicon wafer; a laser centered at 532 nm was used as the light source, and all spectra were obtained with 12 s and 10 acquisitions in the 200–1750 $cm^{-1}$ range.

To verify the reproducibility and homogeneity of the films, the Raman spectra of each film were measured at five independent spots. The films free of the BB analyte did not display bands in the studied spectral region.

TEM images of the Ag-Seeds and prepared Ag-Nplate suspensions were acquired using a Tecnai F20 G$^2$ microscope operated at 200 kV. Then, a few microliters of the diluted suspension were dropped onto an ultrathin lacey carbon-coated copper grid and dried at room temperature.

**Synthesis of Ag-Nplate**

Ag-Nplate were synthesized following a two-step procedure [39]. First, 50 mL of trisodium citrate (2.5 mM), 2.5 mL of polyvinylpyrrolidone (PVP, 500 mg L$^{-1}$) and 400 μL of ice-cold NaBH$_4$ solution (10 mM) were placed in an Erlenmeyer flask and continuously stirred. Immediately thereafter, 50 mL of AgNO$_3$ (0.5 mM) was added at 2 mL min$^{-1}$. The solution changed from uncolored to yellow during the reaction, indicating the formation of Ag-Seeds. In the second step, 50 mL of ultrapure water, 750 μL of ascorbic acid (10 mM), and different quantities of the Ag-Seed solution were mixed and maintained under continuous stirring. Then, 30 mL of AgNO$_3$ solution (0.5 mM) was added to the mixture at a rate of 1 mL min$^{-1}$. Finally, 5 mL trisodium citrate solution (25 mM) was added. The color of the solution varied depending on the quantity of Ag-Seeds employed (Figure 1 (a)).

**SERS Samples Preparation**

A batch of Ag-Nplate with LSPR centered at 815 nm was chosen to investigate their capacity to induce a SERS effect on the detection of brilliant blue FCF.

Preparation of the substrate: A 1 w/v % agarose matrix was used as the substrate. First, a 1 w/v % agarose solution was prepared by heating an adequate quantity of agarose in ultrapure water (for control experiments without Ag-Nplate) or in Ag-Nplate suspension. When the agarose was dissolved, 400 µL of the hot solution was placed in a square template (1 cm²) and left to cool and jellify. When jellified, 100 µL of different concentrations of BB ($1\times10^{-2}$, $1\times10^{-3}$, $1\times10^{-4}$ and $1\times10^{-5}$ M) were added and left to diffuse on 400 µL of agarose gel. After these dilutions in the agarose matrix, the concentrations were $2\times10^{-3}$, $2\times10^{-4}$, $2\times10^{-5}$ and $2\times10^{-6}$ M. Afterwards, the substrates were dried at room temperature and in the dark for five days (films with $1\times10^{-6}$, $1\times10^{-7}$, $1\times10^{-8}$, and $1\times10^{-9}$ mol cm$^{-2}$ were obtained for each concentration). To measure the UV-Vis spectrum of the matrix containing agarose and nanoparticles (Nparticles), an easier handling 1 cm² substrate was prepared. To prepare the Nplates substrate, the volume of synthesis previously detailed was doubled. Then, the Ag-Nplate suspension was centrifuged, a portion of the supernatant was removed to obtain a final volume of 10 mL, and then 100 mg of agarose was added to obtain a 1 w/v % agarose/Nparticles suspension. Four different setups were prepared: 1 w/v % agarose, 1 w/v % agarose with Ag-Nplate, 1 w/v % agarose with different concentrations of BB, and 1 w/v % agarose with different concentrations of BB and Ag-Nplate.

**Results and Discussion**

**Characterizations of Ag-Nplate**

As mentioned above, using the seed-mediated growth technique, the size of the Ag-Nplate can be controlled by adding different amounts of Ag-Seeds.

Figure 1 (a) shows a photograph of the obtained suspensions. The first cuvette (from left to right) contained Ag-Seeds, while the others were suspensions of the grown Ag-Nplate. The size and LSPR of the Ag-Nplate varied depending on the amount of Ag-Seeds used (600, 500, 475, 450, 400, 350, 300, 250, 150, 75, 50 and 10 µL), it can be observe that for greater amount of Ag-Seed, shorter Ag-Nplate and LSPR focused on shorter wavelengths were obtained (Figure 1 (b)).

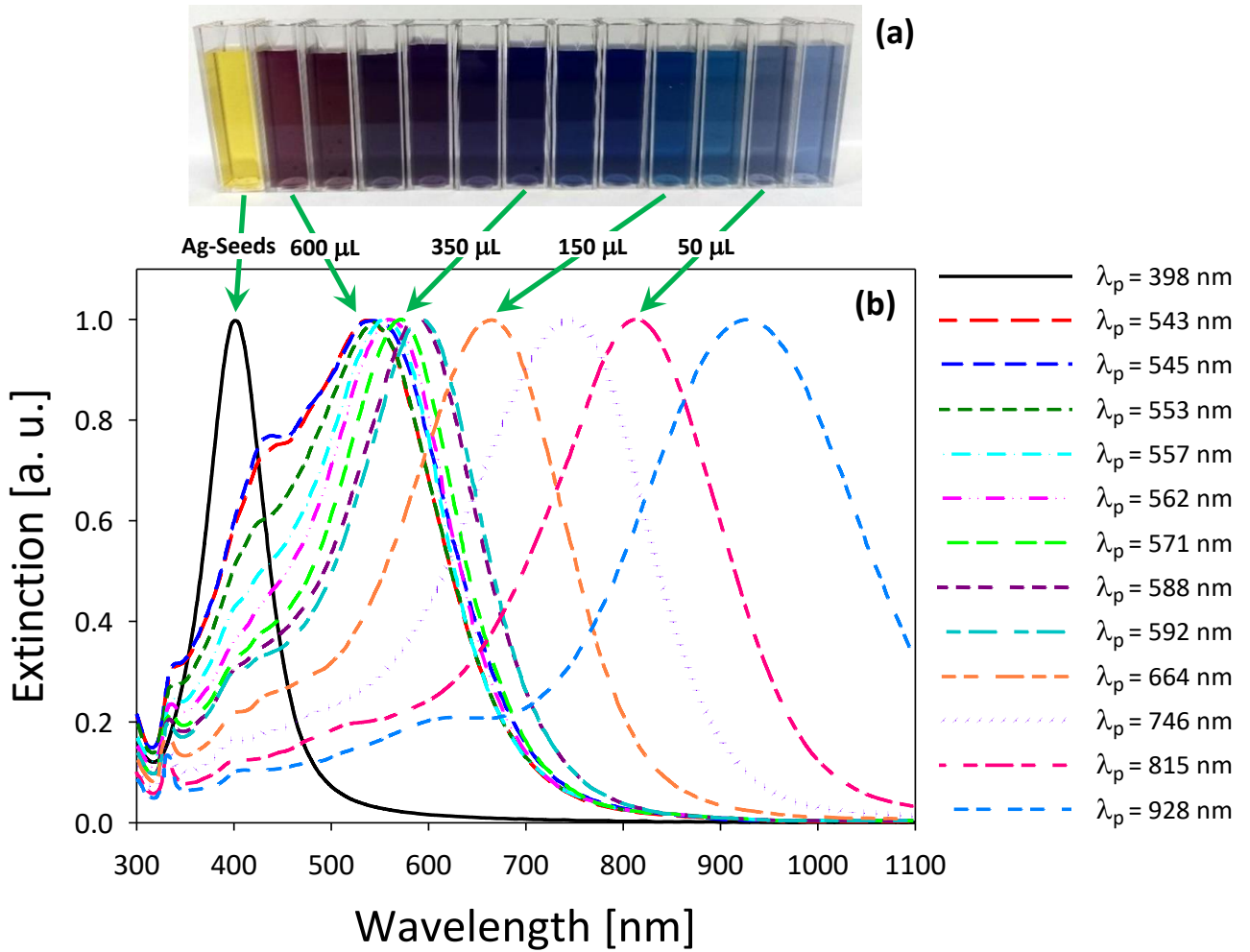

Figure 1. (a) Snapshot of the suspensions obtained. (b) Normalized experimental extinction spectra of the different suspensions.

The size was quantitatively analyzed by theoretical fitting of the experimental extinction spectra. For Ag-Seed suspensions, the fits were performed using Mie theory [51] for metallic Nspheres, while in the case of Ag-Nplates, it was performed using the DDA method [52], considering the experimental bulk dielectric function given by K. M. McPeak et al. [53]. For the case of the Ag-Seeds suspensions, corrective terms for free and bound electron contributions [54], together with the Drude parameters taken from Mendoza-Herrera et al. [55].

Figure 2 (a) shows the normalized experimental extinction spectrum of Ag-Seeds (full line) together with the corresponding fit (dashed-dotted line) calculated using the Mie theory. The optimum log-normal size distribution (inset) is made up of Ag-Seeds with a modal radius of approximately R = 2.22 nm. The size distribution was also obtained from different TEM images of the synthesized suspension. Typical images for Ag Nparticles are shown in Figure 2 (b) and (c). Panel (b) is a low magnification image showing the spherical morphology of the obtained Nparticles, spanning a radius ranging from 1 nm to 5 nm. Panel (c) shows an enlarged TEM image. It can be seen that there are Nparticles (indicated by arrows) with sizes around the modal radius of the size distribution that fits the experimental extinction spectrum. Panel (d) shows the log-normal size distribution (long dashed line) with a modal radius of approximately 2.21 nm, which fits the radius histogram obtained from several images. In this case, the results obtained by TEM were in good agreement with the discussed OES results.

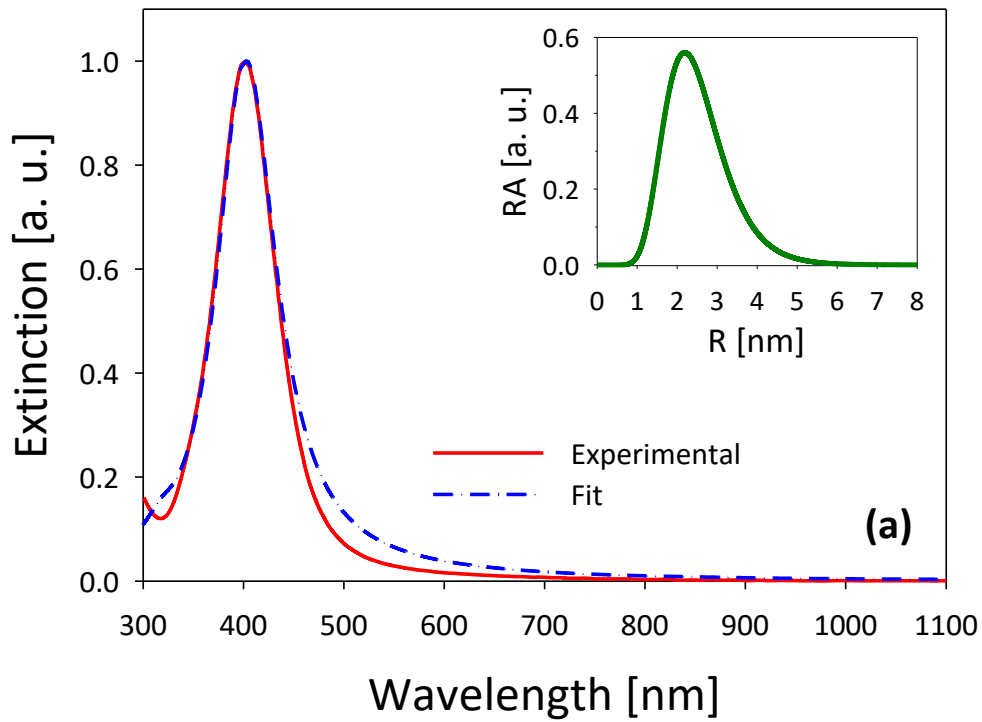

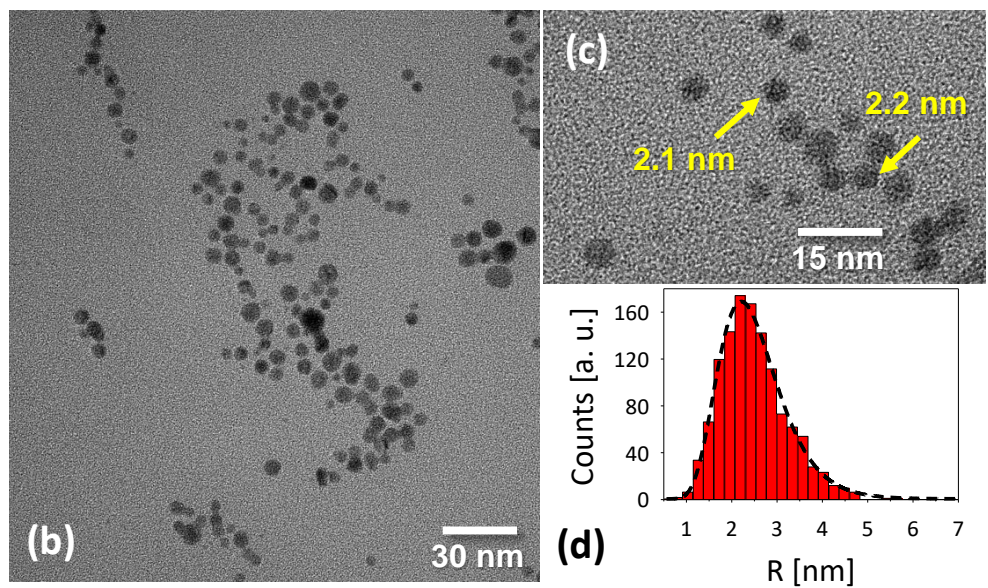

Figure 2. (a) Normalized experimental extinction spectra for freshly prepared Ag-Seeds suspensions with LSPR centered at 398 nm together with theoretical fit. Inset shows the Relative Abundance (RA) of the Ag-Seeds with a size distribution having a modal radius about 2.22 nm. (b) Typical panoramic view of Nparticles TEM image. (c) Enlargement of another TEM image of a group of Ag-Seeds with sizes around the modal radius of the size distribution that fits the experimental extinction spectrum. d) Radius histogram derived from several TEM images together with a log-normal size distribution (long dashed line).

Regarding triangular Nplates extinction spectra, it is not possible to find a full wavelength range fit since the DDA and theory Mie models are based on many parameters (size, edge length and thickness of the Nplates, for the case DDA, and size, bare core and core-shell species of the Nparticles, for the case Mie theory, and bulk dielectric function for both cases), and an absolute minimum in the algorithm cannot be found in a reliable way. However, if the analysis is restrained around the plasmonic resonance region, the shape of the spectra can be approximately modelled considering a triangular contribution and spherical contribution.

Figures 3 and 4 show the normalized experimental extinction spectra (full line) of the suspensions prepared using the method described above with the addition of 600 µL ($\lambda_p$= 543 nm) and 50 µL ($\lambda_p$= 815 nm) of Ag-Seeds, respectively, together with the corresponding total theoretical fits (dashed-dotted lines) calculated using the Mie theory and DDA method. These suspensions contain silver Nspheres (Ag-Nspheres) (SC, long dashed lines) and Ag-Nplate (PC, dashed-double dotted lines) contributions, which are considered for their respective total fits. In Figure 3 (a), the optimum log-normal size distributions (insets) consider a triangular Ag-Nplate contribution (dashed-double dotted line) determined using the DDA method with a modal edge length of approximately $L = 16.30$ nm, and an Ag-Nsphere contribution (long dashed line) calculated using Mie theory with a modal radius of $R = 3.76$ nm. Panel (b) shows a TEM image of a group of Nparticles with a predominantly spherical shape; however, triangular Ag-Nplate with an edge length around 16.20 nm. Panels (c) and (d) show the log-normal size distributions (long dashed line) with a modal radius of approximately 3.20 nm and modal edge length about 15.60 nm, respectively, which fit the radius and edge length histograms obtained from several images. Although TEM showed lower statistics than those obtained by OES, negligible differences were observed in this case.

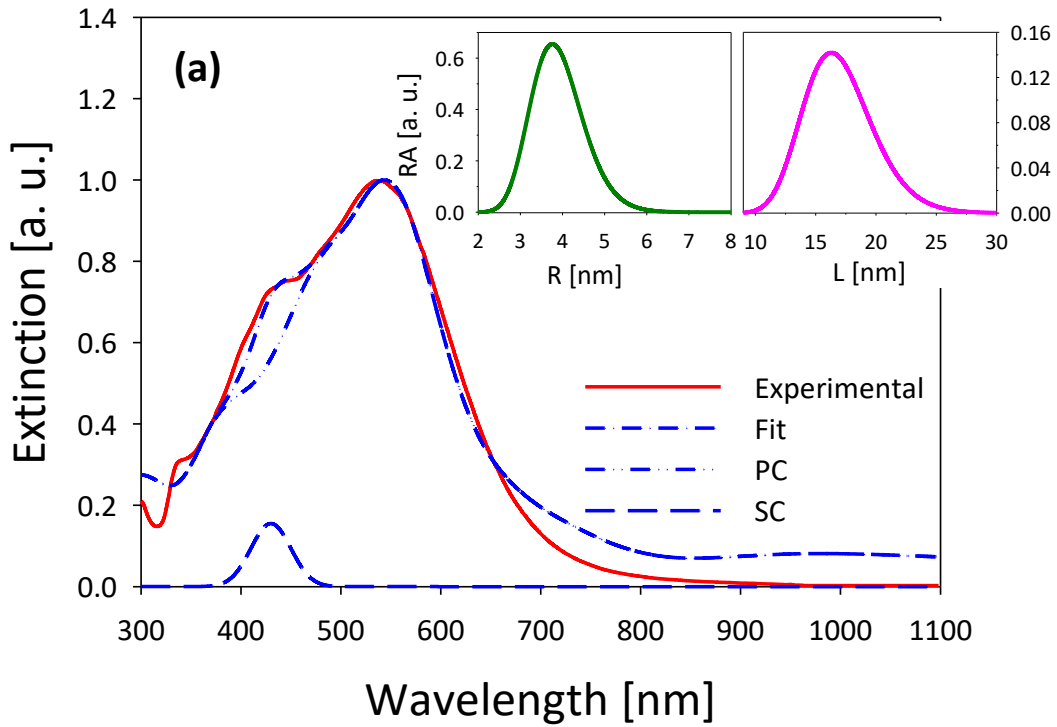

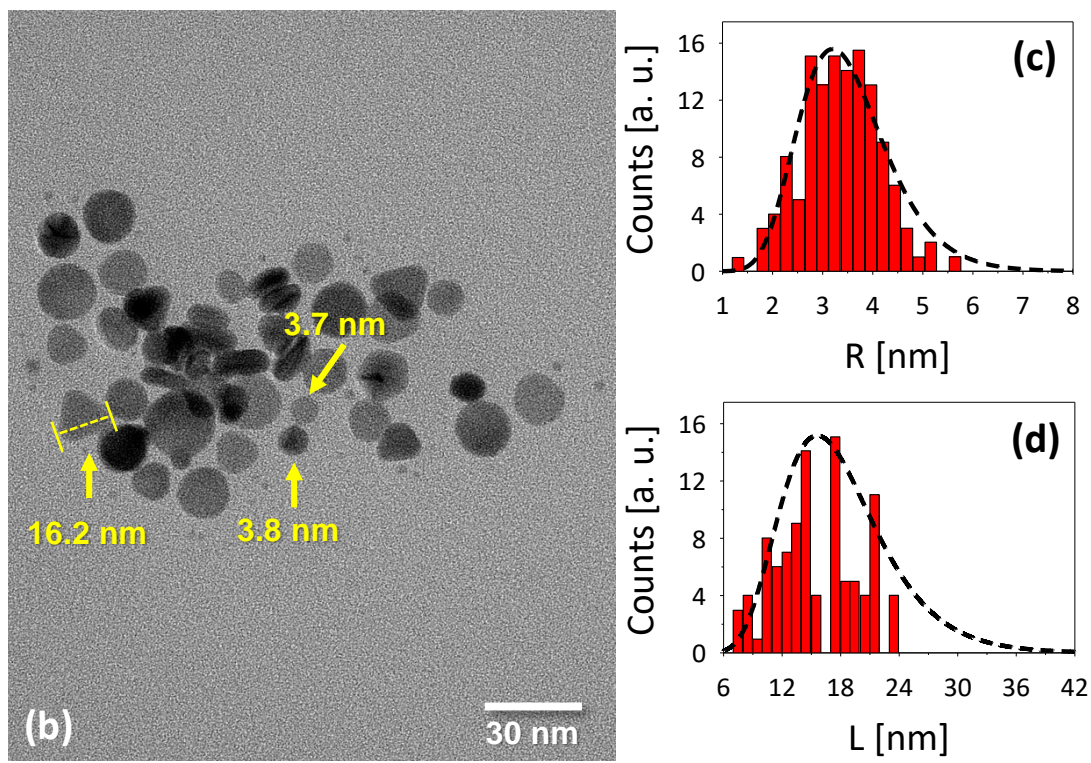

Figure 3. (a) Normalized experimental extinction spectrum for freshly prepared Ag-Nplate suspension with LSPR centered at 543 nm together with theoretical fit. Inset shows the RA of size distributions having an Ag-Nspheres contribution with a modal radius about 3.76 nm and an Ag-Nplate contribution with a modal edge length about 16.30 nm. (b) TEM imagen of a group of Nparticles with sizes around the modal radius and modal edge length. (c) and (d) radius and edge length histograms taken from several TEM images together with their respective log-normal size distributions (long dashed line).

For the case of the Ag-Nplate suspension with a plasmon peak around 815 nm (Figure 4 (a)), the optimum log-normal size distributions (insets) considered an Ag-Nplate contribution (dashed-double dotted line) with a modal edge length of approximately L = 33.82 nm, and an Ag-Nsphere contribution (long dashed line) with a modal radius around R = 7.26 nm. Panel (b) is a panoramic-view TEM image of a group of Nparticles in the presence of triangular Ag-Nplates and Ag-Nspheres. Panel (c) shows an enlarged TEM image. It can be observed that there are Nspheres and triangular Nplates (indicated by arrows) with sizes around the modal radius and modal edge length of the size distributions that fit the experimental extinction spectrum. Panels (d) and (e) show the log-normal size distributions (long dashed line) with a modal radius of approximately 5.40 nm and modal edge length of 33.00 nm, respectively, which fit the radius and edge length histograms obtained from several images. Again, this lower value of the modal radius compared to that obtained by OES may be due to the fact that TEM microscopy has lower sampling statistics.

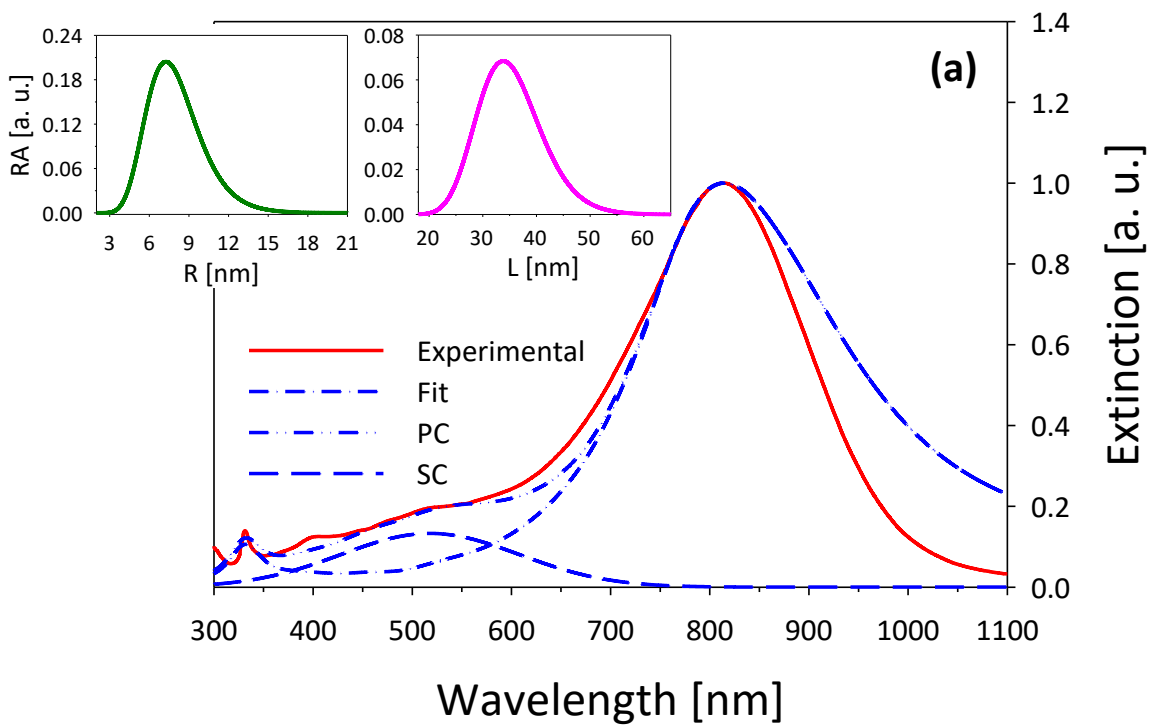

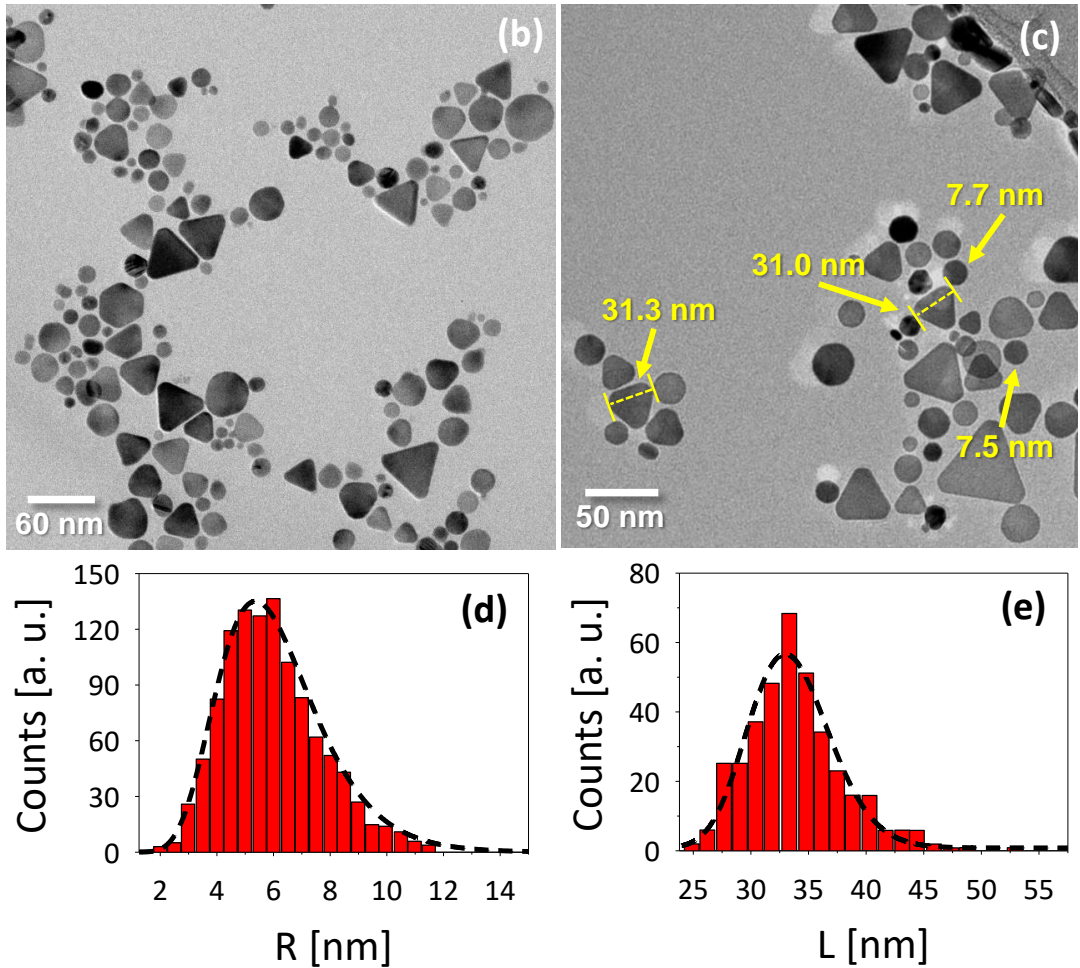

Figure 4. (a) Normalized experimental extinction spectrum for freshly prepared Ag-Nplate suspension with LSPR centered at 815 nm together with theoretical fit. Inset shows the RA of size distributions having an Ag-Nsphere contribution with a modal radius about 7.26 nm and an Ag-Nplate contribution with a modal edge length about 33.82 nm. (b) Panoramic view of a Nparticles TEM image. (c) Enlargement of another TEM image of a group of Nparticles with sizes around the modal radius and modal edge length of the size distribution that fits the experimental extinction spectrum. (d) and (e) radius and edge length histograms taken from several TEM images together with their respective log-normal size distributions (long dashed line).

Different initial amounts of Ag-Seeds were used for the synthesis of the triangular Ag-Nplates. These different amounts of Ag-Seeds give rise to different sizes of triangular Ag-Nplates (green circle in Figure 5), which in the case of extinction spectra is reflected in different positions of the plasmonic peak (blue square in Figure 5). Experimental results suggest that when the initial seed amount is small, the triangular Ag-Nplates formed are larger because the initial seeds have more silver added to grow. As the initial amount of seeds grew, the amount of added silver available to the initial seeds decreased, and the triangular Ag-Nplates formed were smaller. However, this process can also be analyzed from the point of view of the evolution of the size of the initial seeds and the seeds that are left in excess after the triangular Ag-Nplates formation process. The results suggest that the formation of triangular Ag-Nplates is divided into three regions, as shown in Figure 5.

In the region of higher initial seed amount, that is, where the synthesis leads to smaller triangular Ag-Nplates, the larger Ag-Nspheres are mostly used. This is because the larger Ag-Nspheres in the seed solution are closer to the sizes of the smaller triangular Ag-Nplates and therefore are the first to decrease in solution. In the results obtained by TEM and OES, it is observed that the population of Ag-Nspheres decreases mostly in the larger ones, that is, the average size at the end of the process shifts to smaller values.

On the other hand, in the region of lower initial amount of seeds, where larger triangular Ag-Nplates are formed, all sizes of spheres are used, with a greater use of the smaller spheres. In this case, in the final suspension, the main contribution of the Ag-Nspheres corresponds to the larger Ag-Nspheres, although in small amounts compared to the triangular Ag-Nplates. This result was confirmed by both methods (TEM and OES).

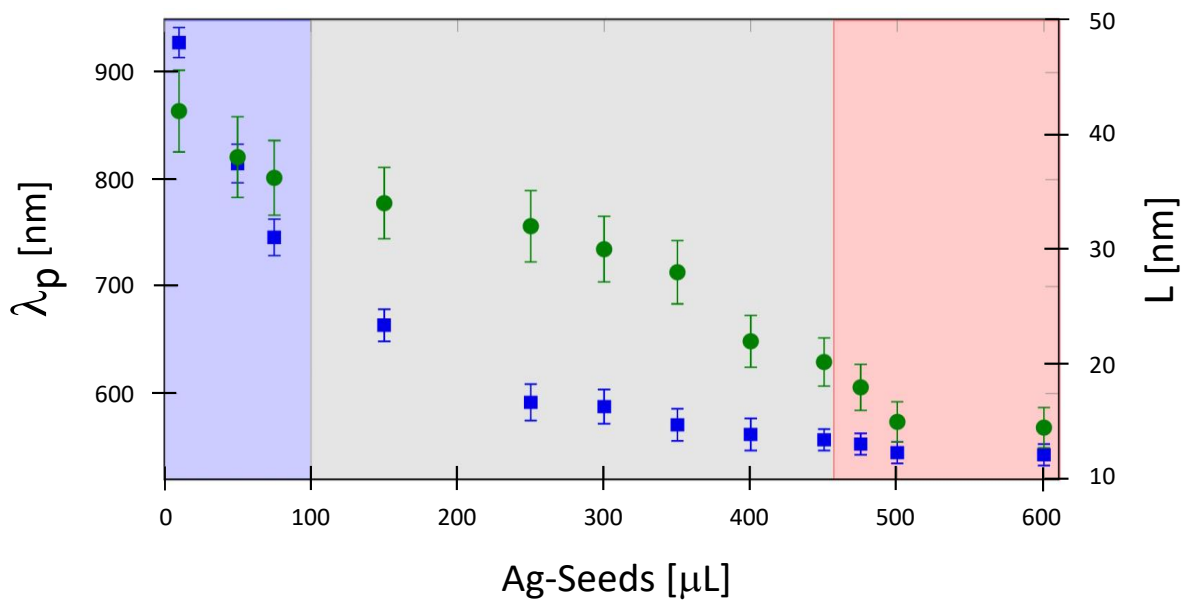

Figure 5. Plasmonic peak position (blue square) and average size (green circles) of the triangular Ag-Nplates as a function of the amount of Ag-Seeds used for the synthesis.

In general, the experimental extinction spectra of the triangular Ag-Nplate suspensions with their respective theoretical results show that the sharp and small peak around 335 nm is attributed to the out-of-plane quadrupole resonance of the triangular Ag-Nplates. The peak near 400 nm may be attributed to the spherical silver Nparticles and the third peak showing a larger wavelength shift corresponds to the plasmonic resonance of the triangular Ag-Nplates.

**SERS experiments**

The preparation of the substrates for SERS experiments required heating the silver nanomaterial suspensions to 100 °C, and the UV-Vis spectra of the different suspensions were recorded before and after heating at 100 °C for 15 min. The stability of the Ag-Nplates in suspension was tested by varying the temperature. The selected samples were studied (LSPR centered at 398, 543, and 815 nm). The absence of spectral changes indicates the stability of the aqueous suspension under these conditions.

The inclusion of the BB dye in the agarose support was performed by diffusion of a solution of the same dye in the gel to rule out possible heterogeneities typical of the diffusion process, and the Raman spectra of each of the supports were determined at least five different points of the same. The tests showed identical spectra, both in the wavenumber of the bands present and in their intensity, which led to the conclusion that homogeneous diffusion of the analyte in the substrate was achieved. The substrates were dried at room temperature in the dark for 5 days. A batch of Ag-Nplates with LSPR centered at 815 nm was used to investigate its ability to induce a SERS effect in the detection of BB.

In Figure 6, the Raman spectra obtained for BB dye in agarose gel with and without Ag-Nplates are shown along with the assignment of selected bands for BB. The inset shows a snapshot of the agarose gel, agarose gel with BB, and agarose gel with BB and Ag-Nplates. The experimental Raman spectrum was consistent with previously reported studies [56].

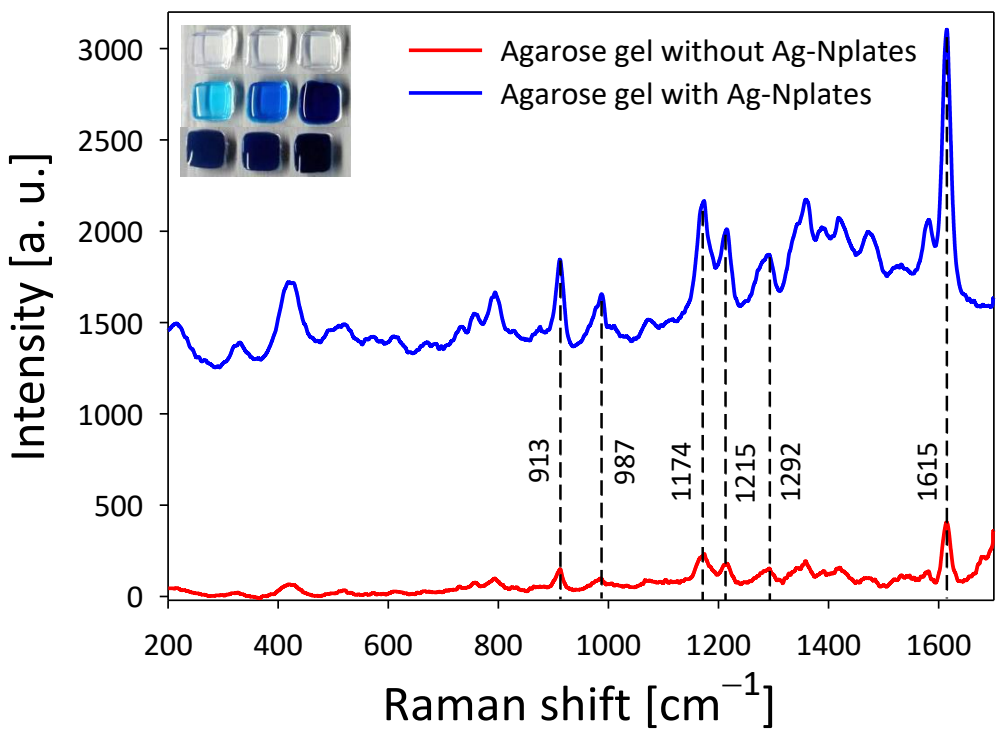

Figure 6. Raman spectra BB dye in agarose gel with Ag-Nplates (blue) and BB dye in agarose gel without Ag-Nplates (red). Inset: agarose gel, agarose gel with BB, and agarose gel with BB and Ag-Nplates.

The Raman spectrum of BB has been previously characterized and assigned based on experimental and DFT calculations [57]. In Table 1, selected bands are listed with their corresponding assignments.

Table 1. Assignment of selected bands of the Raman spectrum of Brilliant Blue.

| Raman Wavenumber [cm$^{-1}$] | Assignment |
|:---:|:---:|
| 1615 | δ C−N; δ C−H |
| 1292 | ν C=C; ω C−N; δ C−H |
| 1215 | ν C=N; ν C=C; ω C−H; δ C−H (ethyl) |
| 1174 | ν C=C; ω C−C; ω C−H |
| 987 | ν C=C; ν S−O |
| 913 | ω C−H (phenyl rings); δ S−O |

Raman spectra of BB supported on agarose and supported on agarose with Ag-Nplates with LSPR centered at 815 nm were analyzed, the assays were performed under the same experimental conditions. The spectrum of BB supported on agarose (without Ag-Nplates) presented bands of low intensity and low resolution. On the other hand, the spectrum of the dye supported on agarose in the presence of Ag-Nplates showed considerably higher intensity and better resolution of the bands. From the numerical integration between 1555 and 1640 cm$^{-1}$ of the Raman spectra, taking five different points of the samples with and without Ag-Nplates, the area under the curve 3.17 times greater in the case of measurements made in the presence of Ag-Nplates was plates.

**Conclusions**

The formation mechanism of triangular Ag-Nplates can be explained by considering the different experimental conditions: in the case where large amounts of seeds are used, smaller triangular Ag-Nplates will form, for which mainly larger Ag-Seeds are used. In the region of lower initial amount of seeds, larger triangular Ag-Nplates are formed, which mostly grow at the expense of smaller spheres. These results were in agreement with those obtained using TEM and OES.

A substrate suitable for the detection of brilliant blue analytes by SERS was obtained. Ag-Nplates with LSPR centered at 815 nm embedded in 1% agarose were used. The Raman spectra obtained in the presence of Ag-Nplates showed higher intensity and better resolution of the bands. The diffusion method used to add the analyte to the supports was satisfactory because the Raman spectra measured at different points on the agarose substrate did not show spectroscopic differences.

**Acknowledgements**

This research was funded by ANPCyT: PICT 2018-3451, PICT APLICADOS 2021-0074 and PICT 2019-02989. C. Villa-Pérez is a CPA member in CONICET, Argentina. J. M. J. Santillán, D. Muñetón Arboleda, L. J. Mendoza-Herrera, and M. S. Moreno are research members at CONICET, Argentina. V. B. Arce is a research member of Comisión de Investigaciones Científicas de la Provincia de Buenos Aires (CIC-PBA).

**Author Declarations**

**Conflicts of Interest**

The authors have no conflicts to disclose.

**Author Contributions**

**Dr. Cristian Villa-Pérez**: silver nanoparticle synthesis (equal), sample preparation (equal), acquisition of optical extinction spectra (equal), conceptualization (equal), investigation (equal), methodology (equal), validation (equal), visualization (equal), and writing - original draft (equal). **Dr. Luis J. Mendoza-Herrera**: theoretical fitting of extinction spectra for silver nanoparticles (equal) and nanoplates (lead), conceptualization (equal), formal analysis (equal), investigation (equal), methodology (equal), shoftware (equal), and writing - original draft (equal). **Dr. Jesica M. J. Santillán**: optical extinction spectra analysis (equal), TEM analysis (equal), conceptualization (equal), formal analysis (equal), investigation (equal), methodology (equal), visualization (equal), writing - original draft (equal), and writing - review & editing (equal). **Dr. David Muñetón Arboleda**: optical extinction spectra analysis (equal), theoretical fitting of nanoparticles extinction spectra (equal), conceptualization (equal), formal analysis (equal), investigation (equal), methodology (equal), software (equal) and acquisition (lead), and writing - original draft (equal). **Dr. M. Sergio Moreno**: TEM analysis supervision (equal) and acquisition (lead), conceptualization (equal), formal analysis (equal), investigation (equal), and writing - original draft (equal). **Dr. Valeria B. Arce**: silver nanoparticle synthesis (equal), sample preparation (equal), acquisition of optical extinction spectra (equal), conceptualization (equal), formal analysis (equal), funding acquisition (lead), investigation (equal), resources (lead), visualization (equal), writing - original draft (equal), and writing - review & editing (equal).

**Data Availability**

The data that support the findings of this study are available from the corresponding author [VBA] upon reasonable request.

**References**

[1] E. L. Hu, S. M. Davis, R. Davis, and E. Scher, *Applications: Catalysis by Nanostructured Materials*, in Nanotechnology Research Directions for Societal Needs in 2020, vol. 1 (Science Policy Reports, Springer, Dordrecht, 2011), pp. 445–466. https://doi.org/10.1007/978-94-007-1168-6_11

[2] L. R. P. Kassab, C. D. S. Bordon, A. S. Reyna, and C. B. de Araújo, "Nanoparticles-based photonic metal–dielectric composites: A survey of recent results," *Optical Materials: X* 12, 100098, (2021). https://doi.org/10.1016/j.omx.2021.100098

[3] H. Lüth, "Semiconductor nanostructures: a new impact on electronics," *Applied Surface Science* 130–132, 855–865, (1998). https://doi.org/10.1016/S0169-4332(98)00166-4

[4] H.-Ch. Kim, S.-M. Park, and W. D. Hinsberg, "Block Copolymer Based Nanostructures: Materials, Processes, and Applications to Electronics," *Chem. Rev.* 110 (1), 146–177, (2010). https://doi.org/10.1021/cr900159v

[5] S. F. Leung, Q. Zhang, F. Xiu, D. Yu, J. C. Ho, D. Li, and Z. Fan, "Light management with nanostructures for optoelectronic devices," *J. Phys. Chem. Lett.* 5 (8), 1479–1495, (2014). https://doi.org/10.1021/jz500306f

[6] A. L. M Reddy, S. R. Gowda, M. M. Shaijumon, and P. M. Ajayan, "Hybrid nanostructures for energy storage applications," *Adv. Mater.* 24 (37), 5045–5064, (2012). https://doi.org/10.1002/adma.201104502

[7] G. Centi, and S. Perathoner, "The role of nanostructure in improving the performance of electrodes for energy storage and conversion," *European Journal of Inorganic Chemistry* 2009 (26), 3851–3878, (2009).

https://doi.org/10.1002/ejic.200900275

[8] N. M. Noah, "Design and synthesis of nanostructured materials for sensor applications," *Journal of Nanomaterials* 2020, article ID 8855321, 20 pp., (2020). https://doi.org/10.1155/2020/8855321

[9] M. Elsabahy, G. S. Heo, S.-M. Lim, G. Sun, and K. L. Wooley, "Polymeric nanostructures for imaging and therapy," *Chem. Rev.* 115 (19), 10967–11011, (2015). https://doi.org/10.1021/acs.chemrev.5b00135

[10] M. M. Bellah, S. M. Christensen, and S. M. Iqbal, "Nanostructures for medical diagnostics," *Journal of Nanomaterials* 2012, article ID 486301, 21 pp., (2012). https://doi.org/10.1155/2012/486301

[11] M. Adabi, M. Naghibzadeh, M. Adabi, M. A. Zarrinfard, S. S. Esnaashari, A. M. Seifalian, R. Faridi-Majidi, H. T. Aiyelabegan, and H. Ghanbari, "Biocompatibility and nanostructured materials: applications in nanomedicine," *Artif. Cells Nanomed. Biotechnol.* 45 (4), 833–842, (2017). https://doi.org/10.1080/21691401.2016.1178134

[12] M. Jahn, S. Patze, I. J. Hidi, I. J., R. Knipper, A. I. Radu, A. Mühlig, S. Yüksel, V. Peksa, K. Weber, T. Mayerhöfer, D. Cialla-May, and J. Popp, "Plasmonic nanostructures for surface enhanced spectroscopic methods," *Analyst* 141 (3), 756–793, (2016). https://doi.org/10.1039/C5AN02057C

[13] O. Kvítek, J. Siegel, V. Hnatowicz, and V. Švorčík, "Noble metal nanostructures influence of structure and environment on their optical properties," *Journal of Nanomaterials* 2013 (1), article ID 743684, 15 pp., (2013). https://doi.org/10.1155/2013/743684

[14] L. F. AL-Badry, "The influence of the nanostructure geometry on the thermoelectric properties," *Physica E: Low-dimensional Systems and Nanostructures* 83, 201–206, (2016). https://doi.org/10.1016/j.physe.2016.05.019

[15] M. Goyal, and M. Singh, "Size and shape dependence of optical properties of nanostructures," *Appl. Phys. A* 126 (3), article number 176, 8 pp., (2020). https://doi.org/10.1007/s00339-020-3327-9

[16] M. B. Mohamed, V. Volkov, S. Link, and M. A. El-Sayed, "The 'lightning' gold nanorods: fluorescence enhancement of over a million compared to the gold metal,"

*Chem. Phys. Lett.* 317 (6), 517–523, (2000). https://doi.org/10.1016/S0009-2614(99)01414-1

[17] Y. S. Yamamoto, and T. Itoh, "Why and how do the shapes of surface-enhanced Raman scattering spectra change? Recent progress from mechanistic studies," *Journal of Raman Spectroscopy* 47 (1), 78–88, (2016). https://doi.org/10.1002/jrs.4874

[18] L. Au, Y. Chen, F. Zhou, P. H. Camargo, B. Lim, Z.-Y. Li, D. S. Ginger, and Y. Xia, "Synthesis and optical properties of cubic gold nanoframes," *Nano Res.* 1, 441–449, (2008). https://doi.org/10.1007/s12274-008-8046-z

[19] C.-H. Kuo, and M. H. Huang, "Facile synthesis of $Cu_2O$ nanocrystals with systematic shape evolution from cubic to octahedral structures," *J. Phys. Chem. C* 112 (47), 18355–18360, (2008). https://doi.org/10.1021/jp8060027

[20] S. Behrens, W. Habicht, K. Wagner, and E. Unger, "Assembly of nanoparticle ring structures based on protein templates," *Adv. Mater.* 18 (3), 284–289, (2006). https://doi.org/10.1002/adma.200501096

[21] Y. Xia, P. Yang, Y. Sun, Y. Wu, B. Mayers, B. Gates, Y. Yin, F. Kim, and H. Yan, "One-dimensional nanostructures: synthesis, characterization, and applications," *Adv. Mater.* 15 (5), 353–389, (2003). https://doi.org/10.1002/adma.200390087

[22] J. Krajczewski, A. Michałowska, and A. Kudelski, "Star-shaped plasmonic nanostructures: New, simply synthetized materials for Raman analysis of surfaces," *Spectrochim. Acta A Mol. Biomol. Spectrosc.* 225, 117469, (2020). https://doi.org/10.1016/j.saa.2019.117469

[23] Y. Zhou, P. Nash, T. Liu, N. Zhao, and S. Zhu, "The large scale synthesis of aligned plate nanostructures," *Sci. Rep.* 6, article number: 29972, (2016). https://doi.org/10.1038/srep29972

[24] S. S. Shankar, S. Bhargava, and M. Sastry, "Synthesis of gold nanospheres and nanotriangles by the Turkevich approach," *J. Nanosci. Nanotech.* 5 (10), 1721–1727, (2005). https://doi.org/10.1166/jnn.2005.192

[25] X. Yu, Z. Wang, H. Cui, X. Wu, W. Chai, J. Wei, Y. Chen, and Z. Zhang, "A review on gold nanotriangles: synthesis, self-assembly and their applications," *Molecules* 27 (24), 8766, (2022). https://doi.org/10.3390/molecules27248766

[26] G. Hu, W. Zhang, Y. Zhong, G. Liang, Q. Chen, and W. Zhang, "The morphology control on the preparation of silver nanotriangles," *Current Applied Physics* 19 (11), 1187–1194, (2019). https://doi.org/10.1016/j.cap.2019.08.002

[27] I. Pastoriza-Santos, and L. M. Liz-Marzán, "Synthesis of Silver Nanoprisms in DMF," *Nano Lett.* 2 (8), 903–905, (2002). https://doi.org/10.1021/nl025638i

[28] Y. Zhang, Y. Luo, J. Tian, A. M. Asiri, A. O. Al-Youbi, and X. Sun, "Rectangular coordination polymer nanoplates: large-scale, rapid synthesis and their application as a fluorescent sensing platform for DNA detection," *PLoS One* 7 (1), e30426, (2012). https://doi.org/10.1371/journal.pone.0030426

[29] J. Xie, J. Y. Lee, D. I. C. Wang, and Y. P. Ting, "Silver nanoplates: from biological to biomimetic synthesis," *ACS Nano* 1 (5), 429–439, (2007). https://doi.org/10.1021/nn7000883

[30] I. Pastoriza-Santos, and L. M. Liz-Marzán, "Colloidal silver nanoplates. State of the art and future challenges," *J. Mater. Chem.* 18 (15), 1724–1737, (2008). https://doi.org/10.1039/B716538B

[31] H. Yang, J. Zhao, D. Li, Y. Cao, F. Li, J. Ma, and P. Liu, "Application of silver nanotriangles as a novel contrast agent in tumor computed tomography imaging," *Nanotechnology* 32 (49), 495705, (2021). 10.1088/1361-6528/ac21ef

[32] A. Amirjani, N. N. Koochak, and D. F. Haghshenas, "Synthesis of silver nanotriangles with tunable edge length: a promising candidate for light harvesting purposes within visible and near–infrared ranges," *Mater. Res. Express*, 6 (3), 036204, (2018). 10.1088/2053-1591/aaf624

[33] R. K. Kannadorai, G. Hegde, and A. Asundi, "Fluorescence enhancement using silver nanotriangle arrays," *J. Nanosci. Nanotechnol.* 12 (5), 3873–3878, (2012). 10.1166/jnn.2012.6143

[34] X. Meng, M. Zhang, L. Liu, J. Du, N. Li, W. Zou, C. Wang, W. Chen, H. Wei, R. Liu, Q. Jia, H. Shao, and Y. Lai, "Rapid and robust analysis of aristolochic acid I in Chinese medicinal herbal preparations by surface-enhanced Raman spectroscopy," *Spectrochimica Acta Part A: Molecular and Biomolecular Spectroscopy* 285, 121880, (2023). https://doi.org/10.1016/j.saa.2022.121880

[35] C. Wang, B. Liu, and X. Dou, "Silver nanotriangles-loaded filter paper for ultrasensitive SERS detection application benefited by interspacing of sharp edges," *Sensors and Actuators B: Chemical* 231, 357–364, (2016). https://doi.org/10.1016/j.snb.2016.03.030

[36] W. A. Murray, J. R. Suckling, and W. L. Barnes, "Overlayers on silver nanotriangles: field confinement and spectral position of localized surface plasmon resonances," *Nano Lett.* 6 (8), 1772–1777, (2006). https://doi.org/10.1021/nl060812e

[37] Q. Zhang, N. Li, J. Goebl, Z. Lu, and Y. Yin, "A Systematic Study of the Synthesis of Silver Nanoplates: Is Citrate a "Magic" Reagent?," *J. Am. Chem. Soc.* 133 (46), 18931–18939, (2011). https://doi.org/10.1021/ja2080345

[38] W. Zhang, G. Hu, W. Zhang, Y. Zhang, J. He, Y. Yuan, L. Zhang, and J. Fei, "A facile strategy for the synthesis of silver nanostructures with different morphologies," *Materials Chemistry and Physics* 235, 121629, (2019). https://doi.org/10.1016/j.matchemphys.2019.05.017

[39] D. Aherne, D. M. Ledwith, M. Gara, and J. M. Kelly, "Optical Properties and Growth Aspects of Silver Nanoprisms Produced by a Highly Reproducible and Rapid Synthesis at Room Temperature," *Adv. Funct. Mater.* 18 (14), 2005–2016, (2008). https://doi.org/10.1002/adfm.200800233

[40] A. J. Haes, J. Zhao, S. Zou, C. S. Own, L. D. Marks, G. C. Schatz, and R. P. Van Duyne, "Solution-phase, triangular Ag nanotriangles fabricated by nanosphere lithography," *J. Phys. Chem. B* 109 (22), 111585–11162, (2005). https://doi.org/10.1021/jp051178g

[41] G. Pasparakis, "Recent developments in the use of gold and silver nanoparticles in biomedicine," *WIREs: Nanomed. Nanobiotechnol.* 14 (5), e1817, (2022). https://doi.org/10.1002/wnan.1817

[42] Y. Yang, J. Murray, J. Haverstick, R. A. Tripp, and Y. Zhao, "Silver nanotriangle array based LSPR sensor for rapid coronavirus detection," *Sensors and Actuators B: Chemical* 359, 131604, (2022). https://doi.org/10.1016/j.snb.2022.131604

[43] R. R. Miranda, I. Sampaio, and V. Zucolotto, "Exploring silver nanoparticles for cancer therapy and diagnosis," *Colloids and Surfaces B: Biointerfaces* 210, 112254, (2022). https://doi.org/10.1016/j.colsurfb.2021.112254

[44] S. C. Boca, M. Potara, A.-M. Gabudean, A. Juhem, P. L. Baldeck, and S. Astilean, "Chitosan-coated triangular silver nanoparticles as a novel class of biocompatible, highly effective photothermal transducers for *in vitro* cancer cell therapy," *Cancer Lett.* 311 (2), 131–140, (2011). https://doi.org/10.1016/j.canlet.2011.06.022

[45] N. Namazi Koochak, E. Rahbarimehr, A. Amirjani, and D. Fatmehsari Haghshenas, "Detection of Cobalt Ion Based on Surface Plasmon Resonance of L-Cysteine Functionalized Silver Nanotriangles," *Plasmonics* 16, 315–322 (2021). https://doi.org/10.1007/s11468-020-01289-2

[46] S. Li, K. Li, X. Li, and Z. Chen, "Colorimetric electronic tongue for rapid discrimination of antioxidants based on the oxidation etching of nanotriangular silver by metal ions," *ACS Appl. Mater. Interfaces* 11 (40), 37371–37378, (2019). https://doi.org/10.1021/acsami.9b14522

[47] K. Wang, and J. Li, "Reliable SERS detection of pesticides with a large-scale self-assembled Au@4-MBA@Ag nanoparticle array," *Spectrochimica Acta Part A: Molecular and Biomolecular Spectroscopy* 263, 120218, (2021). https://doi.org/10.1016/j.saa.2021.120218

[48] M.-Q. He, Y. Ai, W. Hu, X. Jia, L. Wu, M. Ding, and Q. Liang, "Dual-functional capping agent-mediated transformation of silver nanotriangles to silver nanoclusters for dual-mode biosensing," *Anal. Chem.* 95 (14), 6130–6137, (2023). https://doi.org/10.1021/acs.analchem.3c00426

[49] X. Chong, Y. Zhang, E. Li, K.-J. Kim, P. R. Ohodnicki, C.-H. Chang, and A. X. Wang, "Surface-enhanced infrared absorption: pushing the frontier for on-chip gas sensing," *ACS Sens.* 3 (1), 230–238, (2018). https://doi.org/10.1021/acssensors.7b00891

[50] R. K. Sharma, S. Yadav, S. Dutta, H. B. Kale, I. R. Warkad, R. Zbořil, R. S. Varma, and M. B. Gawande, "Silver nanomaterials: synthesis and (electro/photo) catalytic applications," *Chem. Soc. Rev.* 50 (20), 11293–11380, (2021). https://doi.org/10.1039/D0CS00912A

[51] C. F. Bohren, and D. R. Huffman, *Absorption and Scattering of Light by Small Particles* (Wiley-VCH Vergalg GmbH, Weinheim, Germany, 1998). https://doi.org/10.1002/9783527618156

[52] B. T. Draine, and P. J. Flatau, "Discrete-Dipole Approximation For Scattering Calculations," *J. Opt. Soc. Am. A* 11 (4), 1491–1499 (1994). https://doi.org/10.1364/JOSAA.11.001491

[53] K. M. McPeak, S. V. Jayanti, S. J. Kress, S. Meyer, S. Iotti, A. Rossinelli, and D. J. Norris, "Plasmonic Films Can Easily Be Better: Rules and Recipes," *ACS Photonics* 2 (3), 326–333, (2015). https://doi.org/10.1021/ph5004237

[54] D. Muñetón Arboleda, J. M. J. Santillán, L. J. Mendoza Herrera, D. Muraca, D. C. Schinca, and L. B. Scaffardi, "Size-Dependent Complex Dielectric Function of Ni, Mo,

W, Pb, Zn and Na Nanoparticles. Application to Sizing," *J. Phys. D: Appl. Phys.* 49 (7), 075302, (2016). 10.1088/0022-3727/49/7/075302

[55] L. J. Mendoza-Herrera, M. C. Tebaldi, L. B. Scaffardi, and D. C. Schinca, "Determination of thickness-dependent damping constant and plasma frequency for ultrathin Ag and Au films: nanoscale dielectric function," *Phys. Chem. Chem. Phys.* 24, 28019–28028, (2022). https://doi.org/10.1039/D2CP04286J

[56] J. C. Gukowsky, T. Xie, S. Gao, Y. Qu, and L. He, "Rapid identification of artificial and natural food colorants with surface enhanced Raman spectroscopy," *Food Control* 92, 267–275, (2018). https://doi.org/10.1016/j.foodcont.2018.04.058

[57] Y. Xie, T. Chen, Y. Guo, Y. Cheng, H. Qian, and W. Yao, "Rapid SERS detection of acid orange II and brilliant blue in food by using $Fe_3O_4$@Au core–shell substrate," *Food Chemistry* 270, 173–180, (2019). https://doi.org/10.1016/j.foodchem.2018.07.065